\documentclass[journal,twocolumn]{IEEEtran}
\usepackage{epsfig,makeidx,color}
\usepackage{amsmath,amssymb,bbm}
\usepackage{cite,graphicx,lipsum}
\usepackage{enumerate}
\usepackage[switch,pagewise]{lineno}
\usepackage{hyperref}
\hypersetup{
        colorlinks = true,
        citecolor=blue,
}
\def\cQ{{\cal Q}}

\def\rH{{\rm H}}
\def\rT{{\rm T}}

\def\uP{{\mathbb P}}

\def\uE{{\mathbb E}}

\newtheorem{mylemma}{\bf Lemma} 

\def\be{ \begin{equation} }
\def\ee{ \end{equation} }
\def\bea{ \begin{eqnarray} }
\def\eea{ \end{eqnarray} }

\def\by{{\bf y}}

\def\bs{{\bf s}}
\def\ba{{\bf a}}
\def\br{{\bf r}}

\def\bn{{\bf n}}

\def\bw{{\bf w}}

\def\bI{{\bf I}}

\def\bR{{\bf R}}

\def\b0{{\bf 0}}

\def\cC{{\cal C}}
\def\cL{{\cal L}}

\def\cQ{{\cal Q}}

\def\cN{{\cal N}}

\def\cU{{\cal U}}

\def\sH{{\sf H}}

\def\sSNR{{\sf SNR}}

\ifCLASSOPTIONonecolumn
  \newcommand{\figwidth}{0.60\columnwidth}
  
\else
  \newcommand{\figwidth}{0.90\columnwidth}
  
\fi

\begin{document}

\title{Repetition-based NOMA
Transmission and Its Outage Probability Analysis}

\author{Jinho Choi\\
\thanks{The author is with
the School of Information Technology,
Deakin University, Geelong, VIC 3220, Australia
(e-mail: jinho.choi@deakin.edu.au).
This research was supported
by the Australian Government through the Australian Research
Council's Discovery Projects funding scheme (DP200100391).}}


\maketitle
\begin{abstract}
In this paper, we discuss a non-orthogonal multiple access (NOMA)
scheme to exploit a high diversity gain using repetition,
namely repetition-based NOMA.
Unlike conventional power-domain NOMA,
all the users can have the same transmit power, but
different number of repetitions.
Thanks to a high diversity gain,
a low outage probability can be achieved without 
instantaneous channel state information (CSI) 
feedback for power allocation.
A closed-form expression 
for an upper-bound on the outage probability
is derived so that the values of key parameters can be decided
to maintain the outage probability below a target value.
We also consider the average error 
probability for finite-length codes.
Simulation results are compared with the derived bounds
and it is shown that the bounds are reasonably tight and can be used
to decide key parameters (e.g., code rates) to 
guarantee target error probabilities.
\end{abstract}


\ifCLASSOPTIONonecolumn
\baselineskip 28pt
\fi

\section{Introduction}

Since non-orthogonal multiple access (NOMA) 
has a higher spectral efficiency than orthogonal multiple access
(OMA),
it has been extensively studied  \cite{Ding_CM} 
\cite{Choi17_ISWCS}, although there
are a number of challenges (e.g., optimal user clustering 
\cite{Liu16} and beamforming \cite{Choi17_JCN} \cite{Seo18}).
In \cite{Liang17} \cite{ChoiJSAC},
the notion of NOMA can be employed in
uncoordinated transmissions such as 
random access for uplink transmissions
in order to improve the throughput,
which is important for massive machine-type communication
(MTC) that provides the connectivity for various 
Internet-of-Things (IoT) applications
\cite{Bockelmann16}.

In this paper, we consider a NOMA scheme that can have
a low outage probability without 
channel state information (CSI) feedback 
so that it can be used for low-latency communication in MTC
(thanks to no CSI feedback as well as a low outage probability).
In conventional power-domain NOMA,
the power allocation based on instantaneous CSI
is essential for successful successive interference cancellation
(SIC). If instantaneous CSI is not available, 
the power allocation can be carried out with 
statistical CSI to maximize the throughput
as in \cite{Zhang16} (in \cite{Choi17_CSI} for downlink).
In this case, there might be a delay due CSI feedback, and
outage events (due to fading) and error propagation in SIC
are inevitable. As a result, for reliable transmissions,
re-transmissions are required 
and the resulting access delay becomes random and can be long.
Furthermore, as in \cite{Xu16}, when
instantaneous CSI can be available using limited CSI feedback,
due to quantization error and delay \cite{Choi02},
CSI becomes imperfect, which leads to outage events 
and error propagation in SIC.
Thus, if power-domain NOMA is applied to low-latency
communication under time-varying fading
(with statistical CSI or limited CSI feedback),
there might be outage events. 
To keep the outage probability low without instantaneous CSI-based
power allocation, we consider repetition-based NOMA
that exploits a high diversity gain.
The main difference from other power-domain NOMA with statistical
CSI (e.g., \cite{Zhang16} \cite{Choi17_CSI}) is that
the gain from NOMA is used to lower the outage probability
rather than to increase the spectral efficiency.

In order to guarantee a certain low error probability,
good performance prediction techniques are 
required so that the key parameters 
of repetition-based NOMA can be decided in advance.
To this end,
we focus on deriving a tight bound on the outage probability
as a closed-form expression in this paper.
In uplink NOMA, other users' signals become interfering signals
and their strength depends on fading, which makes
the interference power a random variable.
Since its probability density function (pdf)
is unknown, we consider an approximation with
the chi-squared distribution under Rayleigh fading
by using a moment matching approach.
Based on this approximation, we obtain an upper-bound
on the outage probability with finite-length codes
\cite{Polyanskiy10IT}.
We also study decoding error probability 
for finite-length codes using the derived outage probability.
Simulation results show that
this bound is tight at a low outage probability.
Thus, key parameters
(e.g., the code rates) can be determined to keep 
a low outage probability 
or a high probability of successful SIC.

In summary, the main contributions are two-fold:
{\it i)} repetition-based NOMA is proposed to 
exploit a high diversity gain 
for a low error probability without instantaneous 
CSI-based power allocation for low-latency communication;
{\it ii)} a closed-form expression for the outage probability
is derived, which allows to decide key parameters of 
repetition-based NOMA for a desirable 
performance (i.e., a certain low error rate).



The rest of the paper is organized as follows.
In Section~\ref{S:SM}, we present the system model
for repetition-based NOMA to be used for uplink transmissions.
To see the performance of 
repetition-based NOMA in terms of key parameters, we analyze
the performance and find a tight upper-bound on the outage probability
in Section~\ref{S:PA}. The error probability with
finite-length codes is studied in Section~\ref{S:EP}.
Simulation results are presented in Section~\ref{S:Sim}
together with the bounds, which
can help determine design criteria for 
repetition-based NOMA.
The paper is concluded with some remarks in Section~\ref{S:Conc}.

\subsubsection*{Notation} Matrices and 
vectors are denoted by upper- and lower-case
boldface letters, respectively.
The superscripts $\rT$ and $\rH$
denote the transpose and complex conjugate, respectively.
The Kronecker delta is denoted by $\delta_{l,l^\prime}$,
which is 1 if $l = l^\prime$ and 0 otherwise.
$\uE[\cdot]$
and ${\rm Var}(\cdot)$
denote the statistical expectation and variance, respectively.
$\cC\cN(\ba, \bR)$
represents the distribution of
circularly symmetric complex Gaussian (CSCG)
random vectors with mean vector $\ba$ and
covariance matrix $\bR$.
The Q-function is given by
$\cQ(x) = \int_x^\infty \frac{1}{\sqrt{2 \pi} } e^{- \frac{t^2}{2} } dt$.

\section{System Model}	\label{S:SM}

In this section, we consider a NOMA scheme for uplink based on
repetition with multiple radio resource blocks. 
As in \cite{Malak19}, each block can be seen as
a time-frequency resource block and it is assumed that a signal transmitted  
through a block experiences an independent block fading
\cite{TseBook05}.
Unlike conventional power-domain NOMA schemes,
it is assumed that the average receive powers 
of users' signals in the proposed scheme are
the same (to this end, open-loop power control
can be used, where each user can set its 
transmit power based on the average power of the received signal from a 
base station (BS) based on statistical channel 
reciprocity \cite{ChiangBook}). 
Thus, the proposed scheme may be suitable for uplink with users who cannot 
arbitrarily increase the transmit power for power-domain NOMA
and a BS that does not 
perform instantaneous CSI-based power allocation.

\begin{figure}[thb]
\begin{center}
\includegraphics[width=\figwidth]{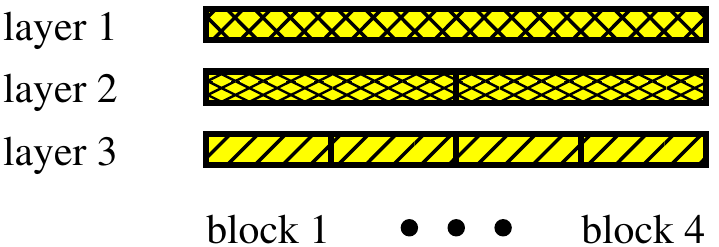}
\end{center}
\caption{A layered structure of frame with 3 layers and 4 blocks.}
        \label{Fig:frame}
\end{figure}

Suppose that we have a set of $L$ parallel radio resource blocks
(in the frequency domain), which is called a frame, for uplink transmissions.
A frame is to be shared by multiple users
in uplink transmissions for a high spectral efficiency based on NOMA. 
For NOMA, there are multiple layers
and a user can transmit a different number of
copies of a packet depending on his/her layer,
where each layer\footnote{In conventional power-domain NOMA, 
each layer is characterized by the transmit power and 
seen as a logical division of a radio resource block for 
superposition coding. In the proposed NOMA scheme, 
each layer is also a logical
division that is characterized by the number of copies.}
is characterized by the number of copies per user
and users in the same layer need to transmit their copies through
different (orthogonal) blocks.
In each block, there are $B$ signals transmitted 
(one signal from each layer),
where $B$ represents the number of layers that are generated
by power-domain NOMA.
For example, as shown in 
Fig.~\ref{Fig:frame}, with a frame consisting of 4 blocks,
we can have 3 layers. A user in layer 1 is to transmit 
4 copies of a packet
through all 4 blocks. There are two users in layer 2 and 
each user is to transmit two copies of a packet through two blocks.
In layer 3, there are 4 users and each user is to transmit
a packet through a block.
In this example, in each frame
consisting of 4 block, there are 7 users in a frame and
3 co-existing signals per block, which clearly
shows that the resulting scheme is a NOMA scheme.
For convenience, the resulting scheme is referred to as repetition-based NOMA.

At the BS, the signals can be decoded with SIC
as other NOMA schemes. In this paper,
it is assumed that a signal in 
a lower layer transmits more copies (repetitions) than
that in a higher layer. Thus, the BS is to decode the signals
from layer 1 to layer $B$ with SIC, where $B$ represents the number of
layers. From this, for successful SIC with a high
probability, it
is expected that the probability of decoding error is the lowest 
in layer 1 and might
increase with $b \in \{1,\ldots, B\}$,
where $b$ represents the layer index,
thanks to the diversity gain\footnote{It will be
shown later that the diversity gain 
can be equal to the number of copies in the presence
of interference under certain conditions.}.
Although there can be as many as layers, 
the number of layers, $B$, has to be limited, while $B$ needs to
be proportional to $L$ for a high spectral efficiency.

Suppose that user $k$ is in layer $1$.
Let $\cU_l$ be the index set of the users transmitting signals 
through the $l$th block.
In addition, denote by $\cL_k$ the index set of the blocks
that are used for multiple transmissions by user $k$.
Then, the received signal at the BS through 
the $l$th block is given by
\be
\br_l = h_{l,k} \bs_{l,k} + 
\sum_{q \in \cU_l \setminus k} h_{l,q} \bs_{l,q} + \bn_l, \ l \in \cL_k,
\ee
where $h_{l,k}$ represents
the channel coefficient from user $k$ to the BS
through the $l$th block,
$\bs_{l,k}$ is the $l$th copy of the signal packet
transmitted (through the $l$th block) by the $k$th user,
and $\bn_l \sim \cC \cN(\b0, N_0 \bI)$ represents the background
noise vector.
For example, in Fig.~\ref{Fig:frame},
with $L = 4$ and $B = 3$, suppose that user 1 lies in layer 1 
and transmits signals through blocks $\cL_1 = \{1,2,3,4\}$.
In addition, let users 2 and 3 be in layer 2
with $\cL_2 = \{1,2\}$ and $\cL_3 = \{3,4\}$, respectively, and
let users 4, 5, 6, and 7 be in layer 3 with
$\cL_4 = \{1\}$, $\cL_5 = \{2\}$, $\cL_6 = \{3\}$,
and $\cL_7 = \{4\}$, respectively.
In this case, with $k = 1$, we have
$\cU_1 \setminus k = \{2, 4\}$, 
$\cU_2 \setminus k = \{2, 5\}$, 
$\cU_3 \setminus k = \{3, 6\}$, and
$\cU_4 \setminus k = \{3, 7\}$. 

It is assumed that the $l$th copy of user $k$'s signal
is an interleaved block of the original signal block, denoted by
$\bs_k$, i.e.,
\be
\bs_{l,k} = \Pi_{l,k}(\bs_k),
\ee
where $\Pi_{l,k} (\cdot)$ denotes
the interleaving operation for the $l$th copy at user $k$.
In particular, a random permutation can be considered for 
the interleaving operation.
In this case, $\Pi_{l,k}$ is seen as a random permutation matrix.
For convenience, let $\Pi^{-1}_{l,k}$ be the deinterleaving operation.
Throughout the paper, we also assume that
$\uE[\bs_k] = \b0$ and $\uE[\bs_k \bs_q^\rH] = P \bI \delta_{k,q}$,
where $P$ represents the signal power of all users.

At the BS, the decoding order with SIC corresponds to the
number of layer, $b$, as mentioned
earlier. That is, the signals in layer 1 are
to be decoded first.
To decode the signal from user $k$, 
the maximal ratio combining (MRC) \cite{BiglieriBook} \cite{ChoiJBook2}
is used with deinterleaved signals as follows:
\begin{align}
\by_k & = \sum_{l \in \cL_k} h_{l,k}^*
\Pi^{-1}_{l,k}(\br_l)\cr
& = \sum_{l \in \cL_k} |h_{l,k}|^2 \bs_k 
+ \sum_{l \in \cL_k} h_{l,k}^* \bw_{l,k},
	\label{EQ:byk}
\end{align}
where
\be
\bw_{l,k} = 
\sum_{q \in \cU_l \setminus k} h_{l,q} 
\Pi_{l,k}^{-1} (\bs_{l,q}) + \Pi_{l,k}^{-1}(\bn_l).
\ee
Once all the signals in layer 1 are decoded,
they can be removed from $\br_l$ using SIC. 
Then, the BS is to decode the signals in layer 2, and so on.
In this case, $\cU_l$ is to be updated 
by removing the indices of the users in layer 1
because their signals in layer 1 are removed.

Decoding can be unsuccessful, which incurs error propagation and
outage events due to incorrect SIC operation \cite{Zhang16}
(which is also true for downlink NOMA as in \cite{Choi17_CSI}).
Thus, it might be important to guarantee
successful decoding with a high probability.
To this end, it is necessary to have a sufficient
number of repetitions or copies
for a high diversity gain.

In repetition-based NOMA, since the powers are fixed
and no power allocation is carried out,
we consider rate allocation for 
successful decoding with a sufficiently high probability.
To this end,  it is necessary to know the error probability
in terms of the code rate and other parameters.
In the next sections, we will focus on the derivation of the 
error rate.

\section{Performance Analysis}	\label{S:PA}

To guarantee a specific target error probability 
and decide key parameters (e.g., the code rate) accordingly at each layer,
we need to predict the performance under given conditions.
To this end, in this section, 
we focus on the performance analysis with the outage probability
to allow such a prediction.

\subsection{Outage Probability}

In this subsection, we assume that 
the number of copies for user $k$ is $D = |\cL_k|$. In addition,
it is assumed that all the interfering signals
in the lower layers are removed by successful SIC.
To find a closed-form expression
for the outage probability,
the following assumptions are mainly considered:
\begin{itemize}
\item[{\bf A1})] The interleaving operation makes
the copies of the signal $\bs_k$ uncorrelated,
i.e.,
\begin{align}
\uE[\Pi_{l,k} (\bs_k) ( \Pi_{l^\prime,q} (\bs_q))^\rH ]& = P \bI 
\delta_{l,l^\prime} 
\delta_{k,q}
\cr
\uE[\Pi_{l,k}^{-1} (\bs_k) ( \Pi_{l^\prime,q}^{-1} (\bs_q))^\rH ]& = P \bI 
\delta_{l,l^\prime}
\delta_{k,q}.
\end{align}
\item[{\bf A2})] The channels are independent Rayleigh fading channels 
with
\be
\uE[h_{l,k} h_{l^\prime, q}^*] = \sigma_h^2 
\delta_{l,l^\prime} \delta_{k,q}.
\ee
Thus, $X_{l,k} = |h_{l,k}|^2$
has the following exponential distribution:
\be
X_{l,k} \sim {\rm Exp}(\sigma_h^2) = \frac{1}{\sigma_h^2} \exp
\left(
- \frac{X_{l,k}}{\sigma_h^2}
\right), \ X_{l,k} \ge 0.
	\label{EQ:hpdf}
\ee
\end{itemize}
Under the assumption of {\bf A1},
we have
\be
\uE[\Pi_{l,k}^{-1} (\bs_{l,q}) 
(\Pi_{l,k}^{-1} (\bs_{l,q^\prime}) )^\rH]
= P \bI \delta_{q, q^\prime}.
\ee
From this, it can be shown that
\begin{align}
\uE[\bw_{l,k} \bw_{l^\prime,k}^\rH ]
= 
\left(\sum_{q \in \cU_l \setminus k} X_{l,q} P \bI
+ N_0 \right) \delta_{l, l^\prime}.
\end{align}
The instantaneous signal-to-interference-plus-noise ratio 
(SINR) for user $k$ in \eqref{EQ:byk} becomes
\begin{align}
\gamma_k 
& = \frac{(\sum_{l=1}^D X_{l,k})^2P}{\sum_{l=1}^D
X_{l,k}(N_0 + P \sum_{q \in \cU_l \setminus k} X_{l,q} )}  \cr
& = \frac{\sum_{l=1}^D X_{l,k}  P}{
\frac{\sum_{l=1}^D (\sum_{q \in \cU_l \setminus k} X_{l,q}) X_{l,k}
}{\sum_{l=1}^D X_{l,k}}P
+N_0 }. 
	\label{EQ:iSINR}
\end{align}

Let $T_k$ denote the SINR threshold for successful decoding.
Then, the outage probability becomes
\be
\uP_k = \Pr(\gamma_k < T_k).
	\label{EQ:pout}
\ee
Thus, we need to find the distribution of the 
instantaneous SINR, $\gamma_k$.

\subsection{SINR Analysis}

For convenience, let
$M = |\cU_l \setminus k|$, i.e., the number of the interfering
signals is denoted by $M$.
If $M = 0$,
under the assumption of {\bf A2} or
from \eqref{EQ:hpdf},
we can show that
\be
\sum_{l=1}^D X_{l,k} = \frac{\sigma_h^2 \chi_{2 D}^2}{2},
\ee
where $\chi_n^2$ represents a chi-squared random
variable with $n$ degrees of freedom.
For convenience, let
$Z_n = \frac{\chi_{2n}^2}{2n}$.
Then, it follows
\begin{align}
\Pr(\gamma_k < T_k)
= \Pr \left( Z_{D} < \frac{T_k}{ D \sSNR} \right),
	\label{EQ:OP_M0}
\end{align}
where
$\sSNR  = \frac{P \sigma_h^2 }{N_0}$
is the signal-to-noise ratio (SNR).
Thus, using the cumulative distribution function (cdf) of 
the chi-squared random variable,
the outage probability can be found.
Furthermore, as shown in Appendix~\ref{A:2},
the following tight upper-bound on the outage probability can be 
obtained:
\begin{align}
\Pr(\gamma_k < T_k) 
\le \frac{1}{D!} \left( \frac{T_k}{\sSNR} \right)^D
e^{- \frac{c_D T_k}{\sSNR}} 
\le\frac{1}{D!} \left( \frac{T_k}{\sSNR} \right)^D,
	\label{EQ:M0}
\end{align}
where
\be
c_D = D e^{-1} \left( D! \right)^{- \frac{1}{D}} \le 1.
	\label{EQ:c_D}
\ee

Unfortunately, if $M > 0$, it is difficult to obtain
a bound on the outage probability. Thus, we have to resort to an 
approximation.
For convenience, let
the interference term in \eqref{EQ:iSINR} be
\begin{align}
P \frac{\sum_{l=1}^D (\sum_{q \in \cU_l \setminus k} X_{l,q}) X_{l,k}
}{\sum_{l=1}^D X_{l,k}}
= \frac{P \sigma_h^2}{2} \sum_{l=1}^D Y_l \alpha_l,
	\label{EQ:A1}
\end{align}
where $\alpha_l 
= \frac{X_{l,k}}{\sum_{l=1}^D X_{l,k}} \ge 0$ 
with $\sum_{l=1}^D \alpha_l = 1$,
and $Y_l = \frac{2}{\sigma_h^2}
\sum_{q \in \cU_l \setminus k} X_{l,q}$.
Clearly, $Y_l$ is a chi-squared random variables with
$2M$ degrees of freedom.
Let $W = \sum_{l=1}^D \alpha_l Y_l$.
If $\alpha_l = \frac{1}{D}$ for all $l$, we can see that
$W$ becomes a scaled chi-squared random
variable with $2DM$ degrees of freedom. However,
since $\alpha_l$ is a random variable,
the resulting approximation may have a lower variance
than the actual one.
Since $W$ can be seen as a weighted sum of chi-squared random
variables, we may consider another chi-squared random
variable to approximate $W$. To this end,
let
\be
\Omega = \frac{\chi_{2N M}^2}{N},
\ee
where $N$ is a parameter to be decided using a moment matching
approach. 
It can be shown that
$\uE[W] = 
\uE\left[ \sum_{l=1}^D \alpha_l Y_l \right] = 2 M$,
because $\uE[\chi_{2n}^2] = 2n$.
In addition, $\uE[\Omega] = 2 M$.
Thus, regardless of the value of $N$, we can see
that $W$ and $\Omega$ have the same mean.
We can consider the 2nd moment and find $N$ such that
\be
\uE[W^2] = \uE[\Omega^2].
	\label{EQ:2nd}
\ee

\begin{mylemma}	\label{L:1}
The value of $N$ satisfying 
\eqref{EQ:2nd} is given by
\be
N = \frac{D+1}{2}.
	\label{EQ:D1}
\ee
\end{mylemma}
\begin{IEEEproof}
See Appendix~\ref{A:1}.
\end{IEEEproof}

In Fig.~\ref{Fig:ex_cdf},
the empirical 
cdf of $W$ is shown with the cdf of $\Omega$ for two different
pairs of $M$ and $D$.
Clearly, thanks to the moment matching (up to the 2nd
moment), the two cdfs become quite similar to each other.

\begin{figure}[thb]
\begin{center}
\includegraphics[width=\figwidth]{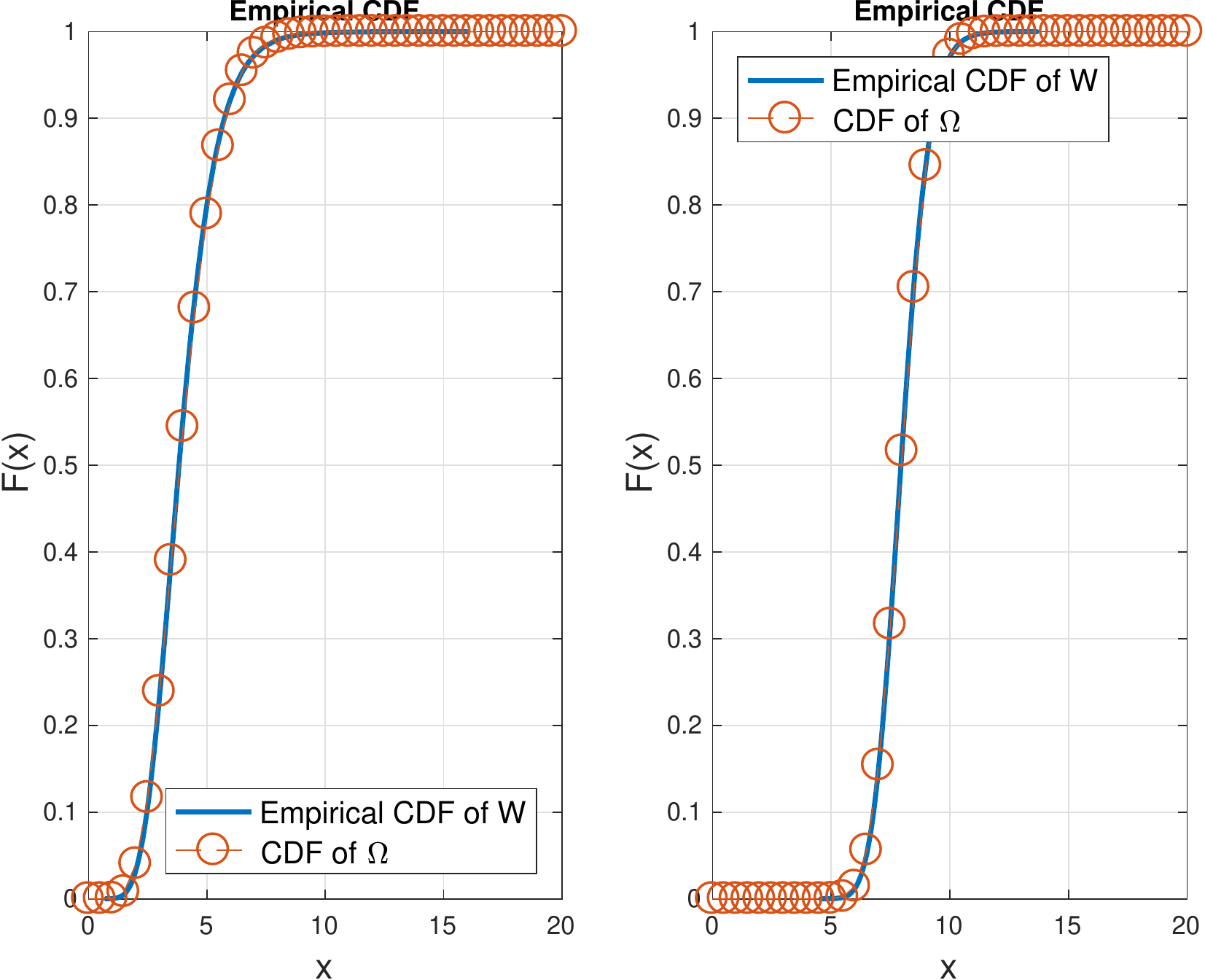}
\end{center}
\caption{Empirical cdf of $W$ and an approximate cdf
with $\Omega$: (a) $(D,M) = (8,2)$; (b) $(D,M) = (32,4)$.}
        \label{Fig:ex_cdf}
\end{figure}

By replacing $W$ with $\Omega$, the
instantaneous SINR in
\eqref{EQ:iSINR} can be
approximated as
\begin{align}
\gamma_k 
\approx \frac{\sum_{l=1}^D X_{l,k} P}{ 
\frac{P \sigma_h^2}{2} \Omega + N_0}.
	\label{EQ:sinrB}
\end{align}
For a tractable analysis, 
the instantaneous SINR in \eqref{EQ:sinrB} can be used.

\subsection{A Closed-form Expression for Outage Probability}

In this subsection, we find a closed-form 
expression for the outage probability with the SINR 
in \eqref{EQ:sinrB}.
Since $\Omega$ is a (scaled) chi-squared random
variable with $2NM$ degrees of freedom,
the outage probability
in \eqref{EQ:pout} can have the following approximation:
\begin{align}
\uP_k 
& = \Pr(\gamma_k < T_k ) \cr
& \approx \tilde \uP_k
= 
\Pr\left( Z_D < \frac{T_k}{D} 
\left( \frac{\chi_{2 N M}^2}{2 N} + \frac{1}{\sSNR}
\right) \right).
	\label{EQ:app}
\end{align}
We can have a closed-form expression for $\tilde \uP_k$
as follows.

\begin{mylemma}	\label{L:2}
For $M \ge 1$, suppose that
\be
d = \frac{D}{c_D T} - \frac{1}{\sSNR} > 0,
	\label{EQ:cond2}
\ee
where $T = T_k$ (for convenience, we omit the index $k$).
Then, we have
\begin{align}
\tilde \uP_k \le \psi (D, M, \sSNR, T) + 
\left( \frac{d e}{M} \right)^{MN} e^{- N d},
	\label{EQ:app2}
\end{align}
where
\begin{align}
& \psi(D,M, \sSNR, T) = 
\frac{1}{D!} \left(\frac{T}{\sSNR} \right)^D
\frac{e^{- \frac{c_D T}{\sSNR}}}{(1+ \frac{c_D T}{N})^{M N}}\cr
& \quad \times
\sum_{n=0}^D \binom{D}{n} 
\left(\frac{\sSNR}{N + c_D T} \right)^n
\prod_{t=0}^{n-1}(M N+t) .
	\label{EQ:psi}
\end{align}
Since the 2nd 
term on the right-hand side (RHS) in \eqref{EQ:app2} is
negligible if $d$ is sufficiently large, 
the first term becomes
a good approximation of $\tilde \uP_k$.
\end{mylemma}
\begin{IEEEproof}
See Appendix~\ref{A:2}.
\end{IEEEproof}

For the outage probability, we will usually consider
the first term on the RHS in \eqref{EQ:app2} (for a large $d$).
Note that in \eqref{EQ:M0}, we can see that
the diversity gain is $D$ 
as the outage probability is proportional to $\sSNR^{-D}$
when $M = 0$.
For the case of $M \ge 1$, in order to see the diversity gain,
we have the following result.

\begin{mylemma}	\label{L:x}
Suppose that \eqref{EQ:cond2} holds. Then, it can be shown that
\be
\psi \le \frac{C}{D!} \nu^D,
	\label{EQ:p_nu}
\ee
where $C$ is a constant that is independent of $D$
and $\nu$ becomes smaller than 1 if
\be
\frac{\sSNR}{T}\left( \frac{D+1}{D+1+2 c_D T} \right)^{\frac{M}{2}}
\ge 1 + \sSNR \frac{ D(M+1) + M-1}{D+1 + 2 c_D T}.
	\label{EQ:condx}
\ee
\end{mylemma}
\begin{IEEEproof}
See Appendix~\ref{A:x}.
\end{IEEEproof}

In \eqref{EQ:p_nu}, taking $\frac{1}{\nu}$ 
as a scaled SINR, we can see that the diversity 
gain\footnote{The diversity order is 
the negative SNR exponent of the outage probability
in a high SNR regime \cite{TseBook05}. In this case, the SNR is replaced
with SINR.}
becomes $D$.
In general, we can derive design criteria
for repetition-based NOMA to keep the outage probability
low from \eqref{EQ:psi} (or to hold \eqref{EQ:condx}). However, 
since the expression in \eqref{EQ:psi} is a bit complicated,
it is not easy to obtain design criteria. 
Thus, for a more tractable analysis,
we can consider the asymptotic $\psi$ when $\sSNR \to \infty$
as follows.

\begin{mylemma}	\label{L:3}
If $M \ge 1$,
we have
\begin{align}
\bar \psi & = \lim_{\sSNR \to \infty} \psi (D, M, \sSNR, T)\cr
& = \binom{MN+D-1}{MN-1}
 \left( \frac{N}{N + c_D T}\right)^{MN}
\left( \frac{T}{N + c_D T} \right)^D.
	\label{EQ:L3}
\end{align}
\end{mylemma}
\begin{IEEEproof}
See Appendix~\ref{A:3}.
\end{IEEEproof}

From \eqref{EQ:L3}, when $D$ increases with a fixed $\rho = \frac{T}{N}$,
since
$\binom{n}{m} \le 2^{n \sH(m/n)}$,
where $\sH(p) = -p \log_2p - (1-p) \log_2 (1-p)$,
we can further show that
\begin{align}
\bar \psi 
& \le 2^{(MN+D-1) \sH \left(
\frac{MN-1}{MN +D -1}
\right)}  
\left( \frac{\rho}{1+ c_D \rho} \right)^D
\left( \frac{1}{1+ c_D \rho} \right)^{MN} \cr
& \approx 
2^{ D \left(1+ \frac{M}{2} \right) \sH \left(\frac{M}{M+2} \right)}
\left(
\frac{\rho}{ \left(1 + \rho\right)^{1 + \frac{M}{2}} }
\right)^D,
\end{align}
where the approximation is tight if $D$ is sufficiently large
with $N = \frac{D+1}{2}$ and $c_D \to 1$.
From this, 
it can be further shown that
\begin{align}
\frac{\log_2 \bar \psi }{D}
& =
\left(1 + \frac{M}{2} \right)
\left(\sH \left(\frac{M}{M+2} \right)
- \log_2 (1 + \rho)
\right) \cr
& \quad + \log_2 \rho.
\end{align}
This implies that if 
\be
\sH \left(\frac{M}{M+2} \right)
< \log_2 (1 + \rho) - \frac{\log_2 \rho}{1 + \frac{M}{2}}
	\label{EQ:cc}
\ee
holds, $\bar \psi$ decreases exponentially with $D$
(i.e., the diversity order
is $D$).
Since $\sH(p) \le 1$, a sufficient condition for \eqref{EQ:cc}
can be found as follows:
\be
2^{1 + \frac{M}{2}} < 
\frac{1}{\rho} =
\frac{N}{T} = \frac{D+1}{2T}.
	\label{EQ:cc2}
\ee

For convenience, let $D_{(b)}$ and $M_{(b)}$
denote the number of copies 
and the number of interfering signals for a user in layer $b$, 
respectively, and assume that all the users in a layer
have the same number of copies and the same number of interfering signals.
Then, according to \eqref{EQ:cc2},
with a sufficiently high SNR,
at layer $b$,
a low outage probability is expected
in repetition-based NOMA if
\be
M_{(b)} \le 2 \log_2 \frac{D_{(b)}}{4 T_{(b)}},
	\label{EQ:MDT}
\ee
where $T_{(b)}$ is the threshold for the users
in layer $b$.
In addition, since $M_{(b)}$ decreases with $b$,
the number of copies in layer $b$,
$L_{(b)}$, needs to be larger than that 
in layer $b^\prime$ if $b < b^\prime$,
as illustrated in Fig.~\ref{Fig:frame}.

In order to determine key parameters,
from \eqref{EQ:MDT}, with a fixed $T_{(b)} = T$ for all layers, we 
can show that 
\be
D_{(b)} = 4 T 2^{ \frac{M_{(b)}}{2}} \propto  2^{ \frac{B-b}{2}},
	\label{EQ:Db}
\ee
since $M_{(b)} = B - b$.
From \eqref{EQ:Db}, we can see that the 
number of copies can decrease exponentially with $b$.
In addition, the number of blocks, $L$, has to be proportional to $2^B$,
which means that the number of layers, $B$, cannot be 
arbitrarily large with respect to a finite $L$.
A small $B$ 
(e.g., $B \le 3$) is also important to keep SIC error propagation limited.
Since $L = D_{(b)} K_{(b)}$, where
$K_{(b)}$ represents the number of users in layer $b$,
we also have $K_{(b)} \propto 2^{\frac{b}{2}}$.

\subsection{Other Issues}

If $B = 1$ (i.e., orthogonal multiple access (OMA) is used), 
each user has the outage probability as in \eqref{EQ:M0}.
However, if we consider $B > 1$ (i.e., NOMA) for a higher spectral
efficiency, the performance of layer $b$
is affected by the performance of layers $1,\ldots, b-1$
through error propagation.
To see the impact of imperfect SIC 
through error propagation on performance,
consider the error probability with SIC propagation.
Let $\epsilon_b$ denote the outage probability of the signal in layer $b$.
Then, the error probability with SIC propagation
at each layer, denoted by $\rho_b$, becomes
\begin{align*}
\rho_1 & = \epsilon_1 \cr
\rho_2 & = (1-\epsilon_1) \epsilon_2 + \epsilon_1 \le \epsilon_1 
+ \epsilon_2 \cr
\rho_3 & = \epsilon_1 + 
(1-\epsilon_1) \epsilon_2 + 
(1-\epsilon_1) (1-\epsilon_2) \epsilon_3  \le 
\epsilon_1 + \epsilon_2 + \epsilon_3 \cr
 & \vdots
\end{align*}
Thus, the error probability with SIC propagation
is bounded by $\sum_{b=1}^B \epsilon_b$ or $O(\max_b
\epsilon_b)$ if $B$ is sufficiently
small (e.g., 3 or 4), which implies that
the impact of error propagation 
on the performance may not be significant.
To see further, suppose that $T_k$ is decided
for a low outage probability,
which is denoted by $\epsilon_{\rm out}$ (and
$\epsilon_{\rm out} = \epsilon_b$ for all $b$),
using the upper-bound on the outage probability 
in \eqref{EQ:app2} for given $D$, $M$, and $\sSNR$.
For example, suppose that $L = 4$ and $B = 3$ as in Fig.~\ref{Fig:frame}.
For user 1 in layer 1, with $D = 4$ and $M = 2$,
$T_1$ can be obtained for a target outage probability, $\epsilon_{\rm out}$.
For user 2 in layer 2,
if $T_2$ is decided to keep a target outage probability
of $\epsilon_{\rm out}$,
the actual outage probability with
taking into account error propagation becomes
\be
\uP_{\rm out} (2) \le 2 \epsilon_{\rm out}.
\ee
Thus, as long as
$\epsilon_{\rm out} \ll 1$ and $B$ is not too large,  
the actual outage probabilities of all the layers can be 
an order of $\epsilon_{\rm out}$ as mentioned earlier.
As a result, it can be seen that 
the proposed repetition-based NOMA transmission
can not only guarantee
a low error probability (without instantaneous CSI-based
resource allocation), but also provide
a high spectral efficiency.

If capacity-achieving codes \cite{MacKayBook} are employed,
the information outage probability  \cite{TseBook05}
can be given by
\be
\Pr(\log_2 (1+ \gamma_k) < R_k) = \Pr(\gamma_k < 2^{R_k} - 1),
\ee
where $R_k$ is the code rate of user $k$'s packet.
Thus, $T_k = 2^{R_k} - 1$.
That is, if $T_k$ is obtained to keep a specific
target outage probability from the closed-form expression
in \eqref{EQ:app2}, 
the corresponding code rate, $R_k$,
can be easily decided.
Therefore,
repetition-based NOMA can guarantee a specific
target error probability (which is usually low enough
to avoid frequent re-transmissions)
without using instantaneous CSI-based power allocation.

\section{Error Probability with Finite-Length
Codes}	\label{S:EP}

In this section, we consider the case
that finite-length codes
are used in repetition-based NOMA.


Suppose that the BS is to decode the signal from user 
$k$ using $\by_k$ in \eqref{EQ:byk}.
For a given $\gamma = \gamma_k$, according to \cite{Polyanskiy10IT}
\cite{Tan15},
the achievable rate
(for complex Gaussian channel \cite{Durisi16})
for a finite-length code
is given by
\be
R^* (n, \epsilon) \approx \log_2 (1 + \gamma)
- \sqrt\frac{ V (\gamma)}{n} \cQ^{-1} (\epsilon) + 
O\left(\frac{\log_2 n}{2 n}
\right),
        \label{EQ:R_PPV}
\ee
where $V(\gamma)$ is the channel dispersion that is given by
$V(\gamma)
= \frac{\gamma(2 + \gamma)}{(1+ \gamma)^2} (\log_2 e)^2$,
$n$ is the length of codeword when a codeword
is transmitted within a block,
and $\epsilon$ is the error probability.
It can be shown that $\bar V > V(\gamma)$, where
$\bar V = \frac{1}{(\ln 2)^2} \approx 2.0814$.
Thus, ignoring the term of
$O\left(\frac{\log_2 n}{n}\right)$
and letting $R = R^*(n,\epsilon)$,
a lower-bound on the achievable rate can be obtained as follows:
\be
R \ge
\log_2 (1 + \gamma)
- \sqrt\frac{\bar V}{n} \cQ^{-1} (\epsilon),
        \label{EQ:DB}
\ee
which might be tight for a sufficiently high SNR, $\gamma$,
because $V(\gamma) \to \bar V$ as $\gamma \to \infty$.
Then, an upper-bound on the average error probability 
is given by
\begin{align}
\bar \epsilon 
& \approx \uE\left[ \cQ \left(
\sqrt{\frac{n}{V (\gamma)}} \left(
\log_2 (1+ \gamma)- R \right) \right) \right] \cr
& \le \uE\left[ \cQ \left(
\sqrt{\frac{n}{\bar V}} \left(
\log_2 (1+ \gamma)- R \right) \right) \right] \cr
& = \frac{1}{\sqrt{2 \pi}} \int_{-\infty}^\infty
\Pr \left(\gamma < 2^{ \sqrt\frac{\bar V}{n} x + R} - 1
\right) e^{- \frac{x^2}{2}} dx,
	\label{EQ:beps}
\end{align}
where the expectation is carried out over $\gamma$
and the first equality is due to \cite[Eq. (3.57)]{VerduBook}.
Note that if $n \to \infty$, $ \sqrt\frac{\bar V}{n} \to 0$.
Thus, we can have
\begin{align*}
\lim_{n \to \infty}
\int_{-\infty}^\infty
\Pr (\gamma < 2^{ \sqrt\frac{\bar V}{n} x + R} - 1
) 
\frac{e^{- \frac{x^2}{2}} }{\sqrt{2 \pi}}  
dx 
= 
\Pr (\gamma < 2^{R} - 1),
\end{align*}
which means that the average error probability becomes
the outage probability.

From \eqref{EQ:beps},
using a closed-form expression for the
outage probability in \eqref{EQ:app2} and a numerical integration
technique, 
we can find an upper-bound on $\bar \epsilon$.


\section{Simulation Results}	\label{S:Sim}

In this section, we present simulation results
that can provide design criteria.
For simulations, we mainly consider the instantaneous
SINR in \eqref{EQ:iSINR} with randomly generated 
channel coefficients according 
to Rayleigh fading channels in \eqref{EQ:hpdf}.

\subsection{Outage Probability}

In this subsection, we present simulation
results of the outage probability.

\begin{figure}[thb]
\begin{center}
\includegraphics[width=\figwidth]{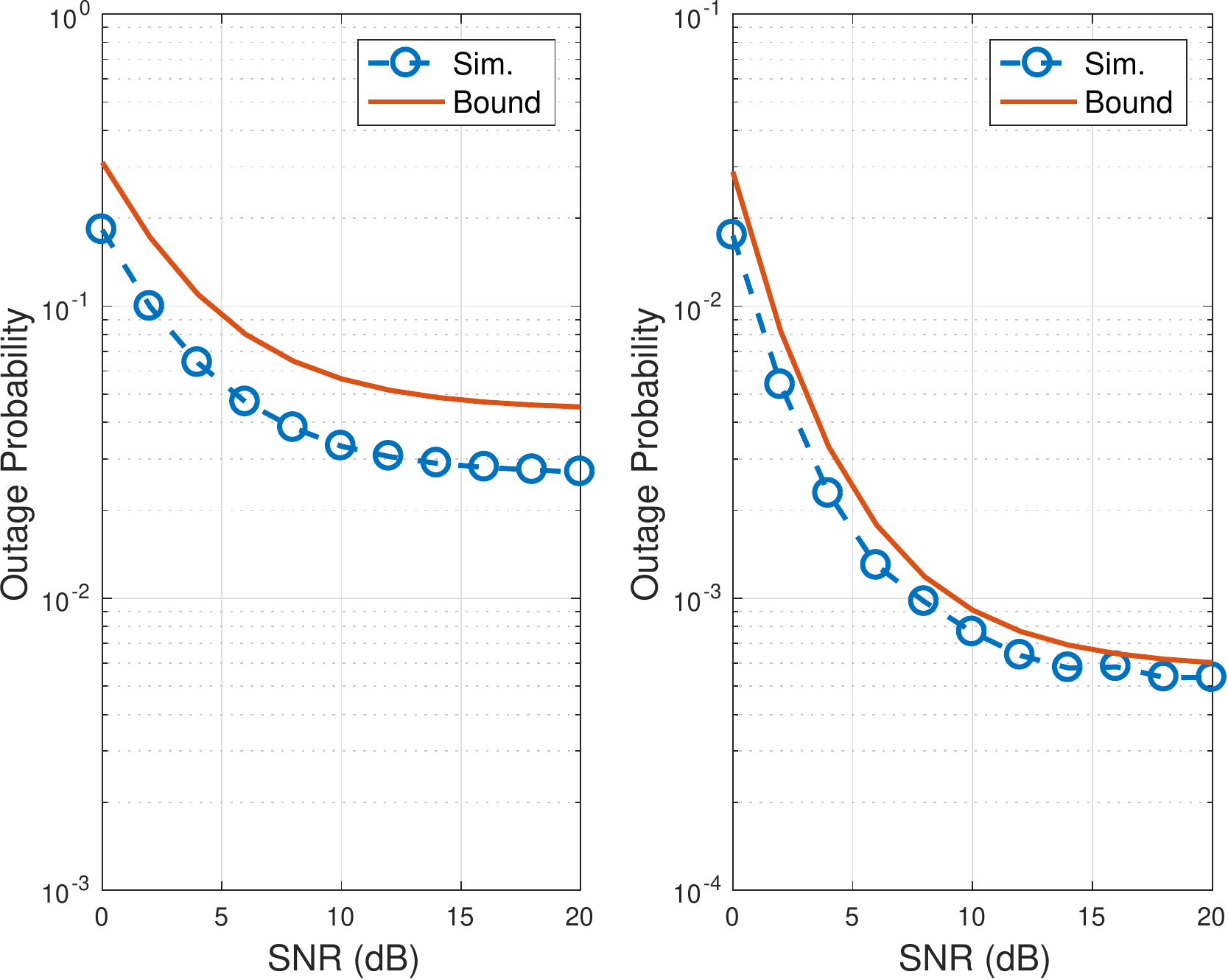} \\
\hskip 0.5cm (a) \hskip 3.5cm (b) 
\end{center}
\caption{Outage probabilities as functions of SNRs
with $D = 16$ and $T = 2$:
(a) $M = 2$; (b) $M = 1$.}
       \label{Fig:plt1}
\end{figure}

Fig.~\ref{Fig:plt1}
shows the outage probability as a function of SNR 
with $D = 16$, $T = 2$, and $M \in \{1,2\}$. 
For the upper-bound, we use the first term 
on the RHS in \eqref{EQ:app2} (in general, the 2nd term is negligible).
Due to the presence of the interfering signals
(as $M \ge 1$), we can see that there is an error
floor in the outage probability.
That is, although the SNR goes to $\infty$,
the outage probability does not approach 0, but
a non-zero constant as shown in \eqref{EQ:L3}.
We can confirm that \eqref{EQ:app2} is a tight upper-bound
from Fig.~\ref{Fig:plt1} (a) and (b).

\begin{figure}[thb]
\begin{center}
\includegraphics[width=\figwidth]{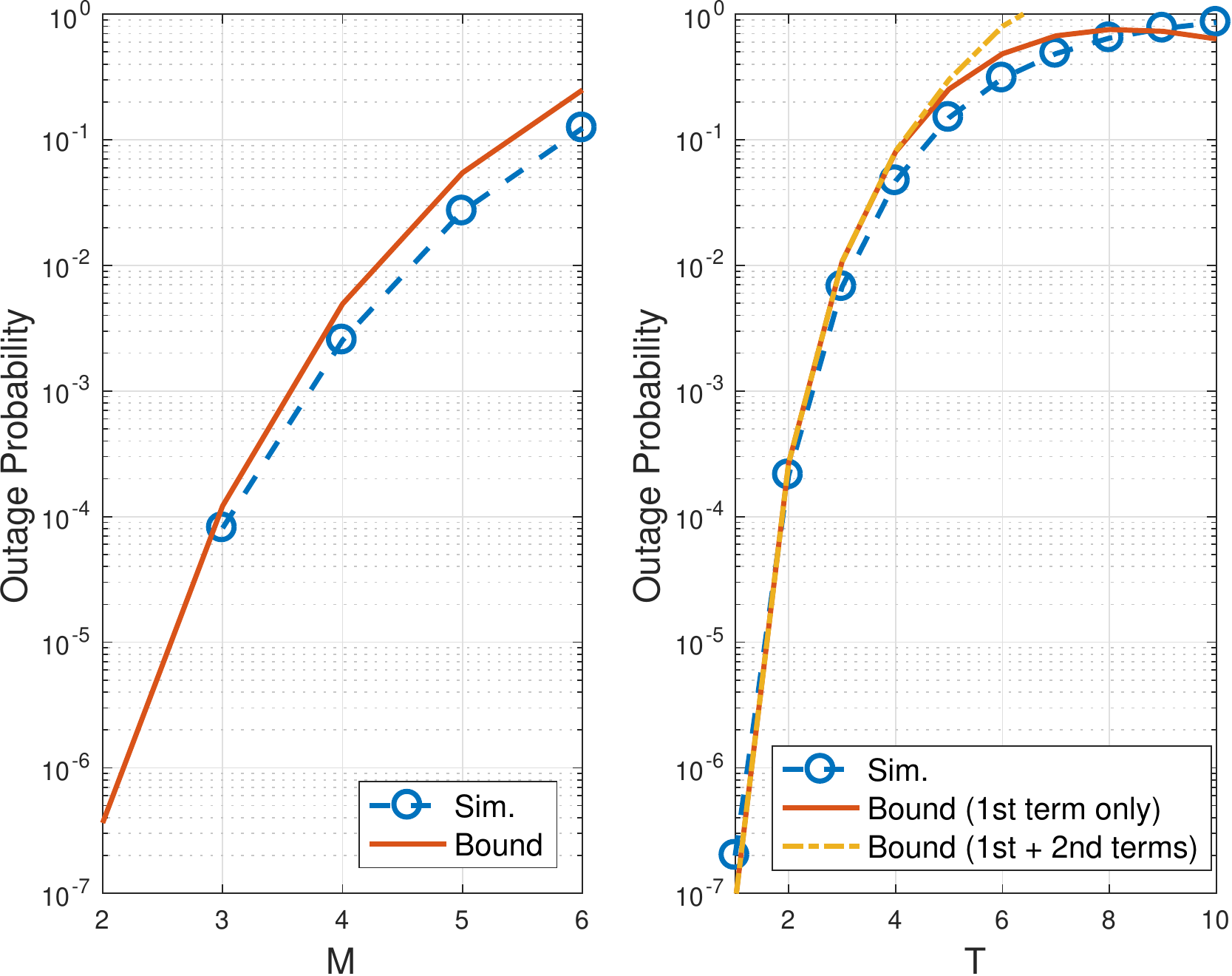} \\
\hskip 0.5cm (a) \hskip 3.5cm (b) 
\end{center}
\caption{Outage probabilities as functions of $M$ and $T$
when $\sSNR = 6$ dB:
(a) outage probability versus $M$ with $D = 32$ and $T = 4$;
(b) outage probability versus $T$ with $D = 16$ and $M = 2$.}
       \label{Fig:plt24}
\end{figure}

The impact of $M$ and $T$ on the outage probability
is shown in Fig.~\ref{Fig:plt24}
when $\sSNR = 6$ dB.
Since $M$ is the number of interfering signals,
the outage probability increases with $M$
as shown in Fig.~\ref{Fig:plt24} (a).
In Fig.~\ref{Fig:plt24} (b),
as expected, the outage probability increases with $T$.
Furthermore, since the bound is tight, we can choose $T$
for a sufficiently low target outage probability.
Note that the first term 
on the RHS in \eqref{EQ:app2} in 
Fig.~\ref{Fig:plt24} (b) is not an upper-bound
when $T$ is large (e.g., $T \ge 8$).
When $T = 10$, the 2nd term on the RHS in \eqref{EQ:app2}
becomes 0.6727, which is not negligible.
As shown in Fig.~\ref{Fig:plt24} (b),
the 2nd term needs to be taken
into account for the upper-bound when $T$ is not small.
However, since we are mainly interested in a low outage probability
(e.g., $\le 10^{-3}$), the impact of the 2nd term on the
upper-bound is negligible.

\begin{figure}[thb]
\begin{center}
\includegraphics[width=\figwidth]{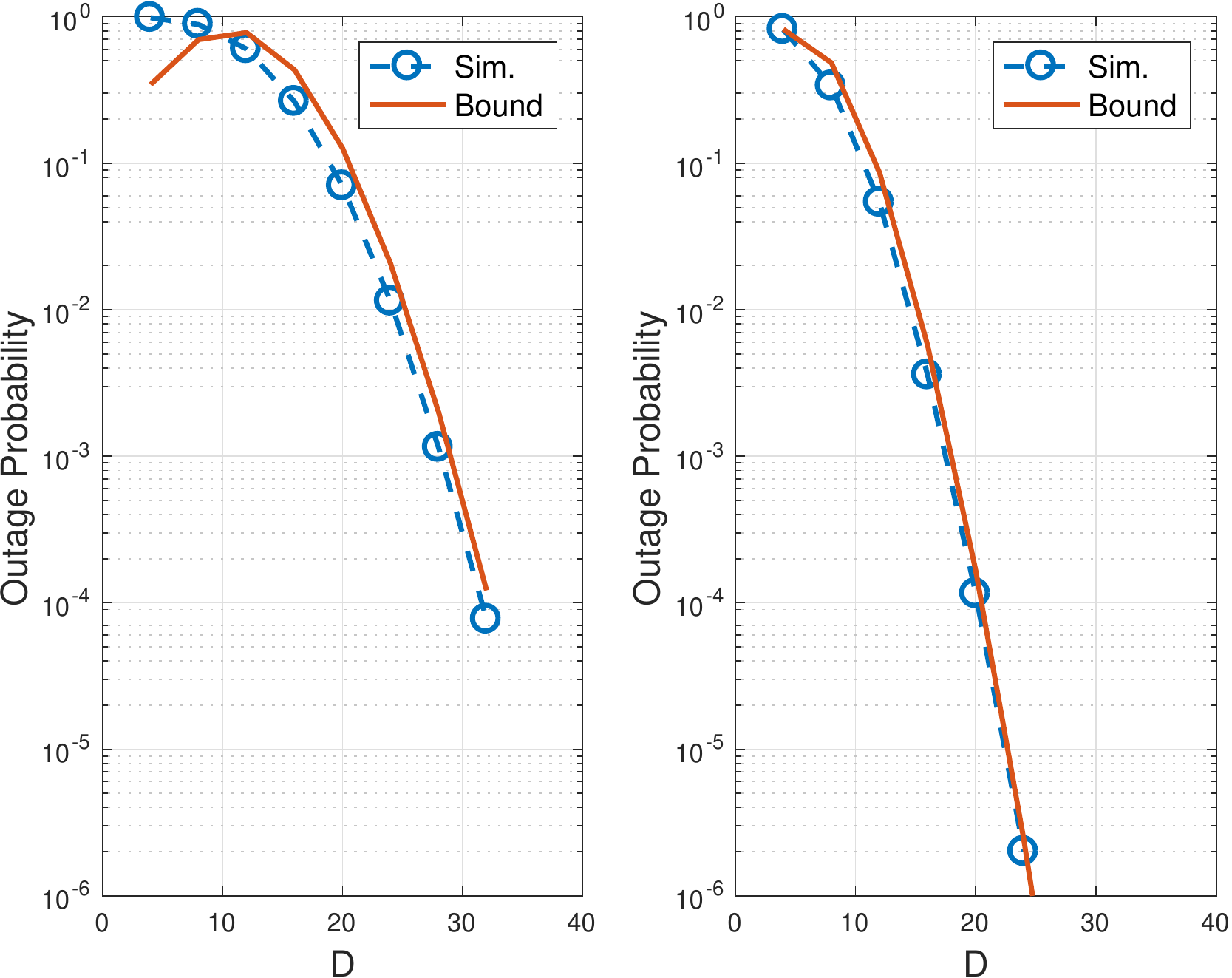} \\
\hskip 0.5cm (a) \hskip 3.5cm (b) 
\end{center}
\caption{Outage probabilities as a function of $D$
with $M = 3$ and $\sSNR = 6$ dB:
(a) $T = 4$; (b) $T = 2$.}
       \label{Fig:plt3}
\end{figure}

Fig.~\ref{Fig:plt3}
shows the outage probability
as a function of $D$
when $M = 3$, $\sSNR = 6$ dB,
and $T \in \{2, 4\}$.
Clearly, a better performance
is achieved with $D$ due to a higher diversity gain.
It is shown that as long as
$d$ is sufficiently large (due to a large $D$ or small $T$),
the bound with the first term in \eqref{EQ:app2}
is reasonably tight and can be used to predict the performance
in terms of the outage probability.

\subsection{Average Error Probability for Finite-Length Codes}

As mentioned earlier, a careful determination of
$T_k$ or $R_k$ is
necessary to keep the outage probability low (for successful
SIC). If finite-length codes are used,
we need to consider the average error probability
instead of the outage probability.
In this subsection, we consider
the average error probability in \eqref{EQ:beps}.

\begin{figure}[thb]
\begin{center}
\includegraphics[width=\figwidth]{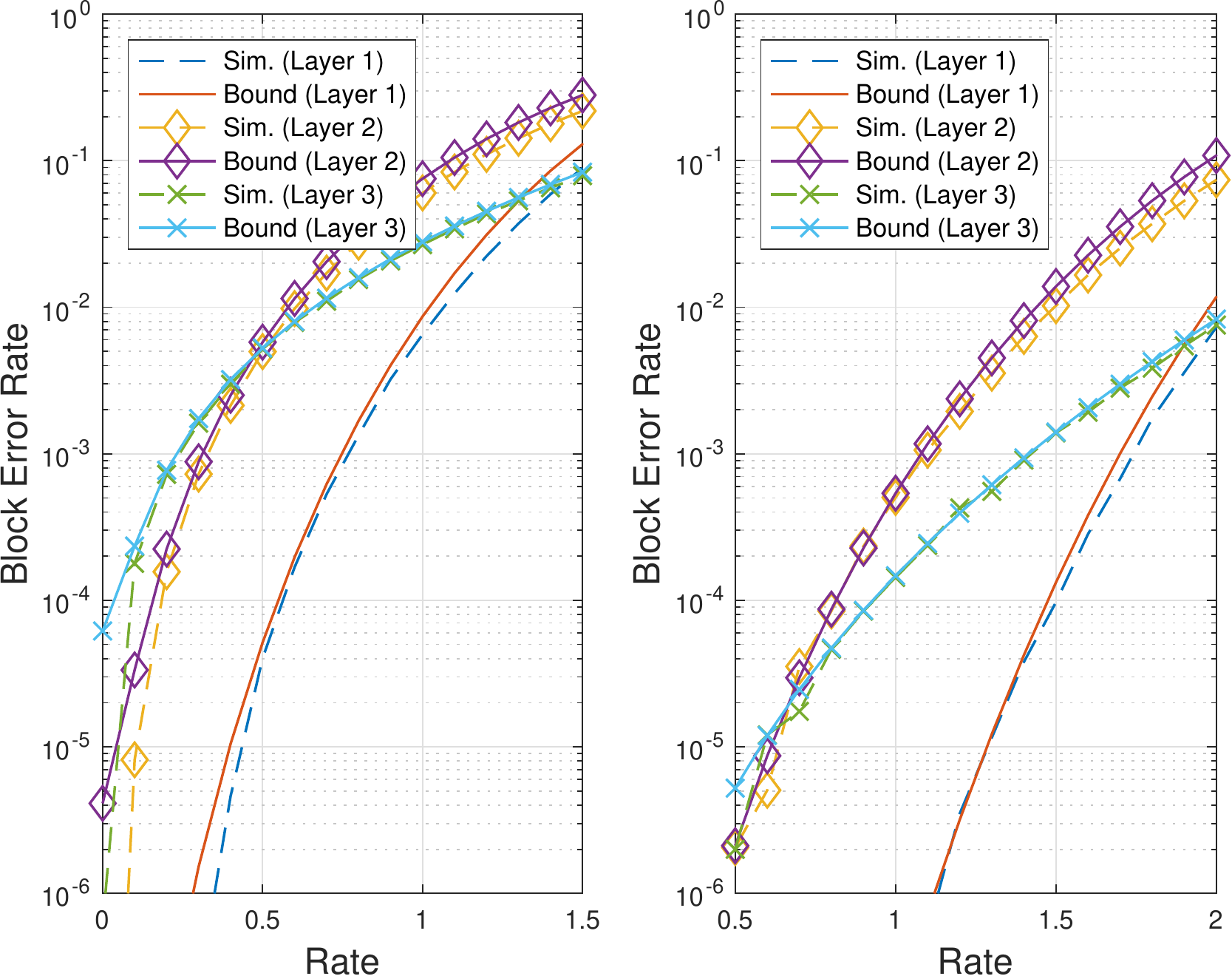} \\
\hskip 0.5cm (a) \hskip 3.5cm (b) 
\end{center}
\caption{Average error probabilities as a function of $R$
with $B = 3$ and $\sSNR = 6$ dB: (a) $L = 8$; (b) $L = 16$.}
       \label{Fig:plt_RvsER}
\end{figure}

In Fig.~\ref{Fig:plt_RvsER},
we consider a repetition-based 
NOMA system with $B = 3$ and $n = 512$ (bits).
It is assumed that one user in layer 1 (transmitting
$L$ copies),
two users in layer 2 (each user 
transmitting $L/2$ copies), and four users in layer 3
(each user transmitting $L/4$ copies).
The average error probability in each layer for 
different values of the code rate, $R$,
is shown in Fig.~\ref{Fig:plt_RvsER}
with the bound from \eqref{EQ:beps}.
As shown in Fig.~\ref{Fig:plt_RvsER} (a),
the signal in layer 1 needs to have $R \le 0.75$ 
for an average error probability
of $10^{-3}$.
It is also possible to decide the code
rates for the signals in 
layers 2 and 3 
for an average error probability
of $10^{-3}$ using the bound from \eqref{EQ:beps},
because the bound is sufficiently tight.
In Fig.~\ref{Fig:plt_RvsER} (b),
we can see that the rates increase more than twice 
when $L = 16$ compared with the rates when $L = 8$
(which are shown in Fig.~\ref{Fig:plt_RvsER} (a))
with a target error probability of $10^{-3}$.
If the target error probability further decreases,
the rate gap increases. 
This indicates that
the repetition-based NOMA scheme
can be more efficient
with a large $L$ and a low target error probability
thanks to a high diversity gain.

\begin{figure}[thb]
\begin{center}
\includegraphics[width=\figwidth]{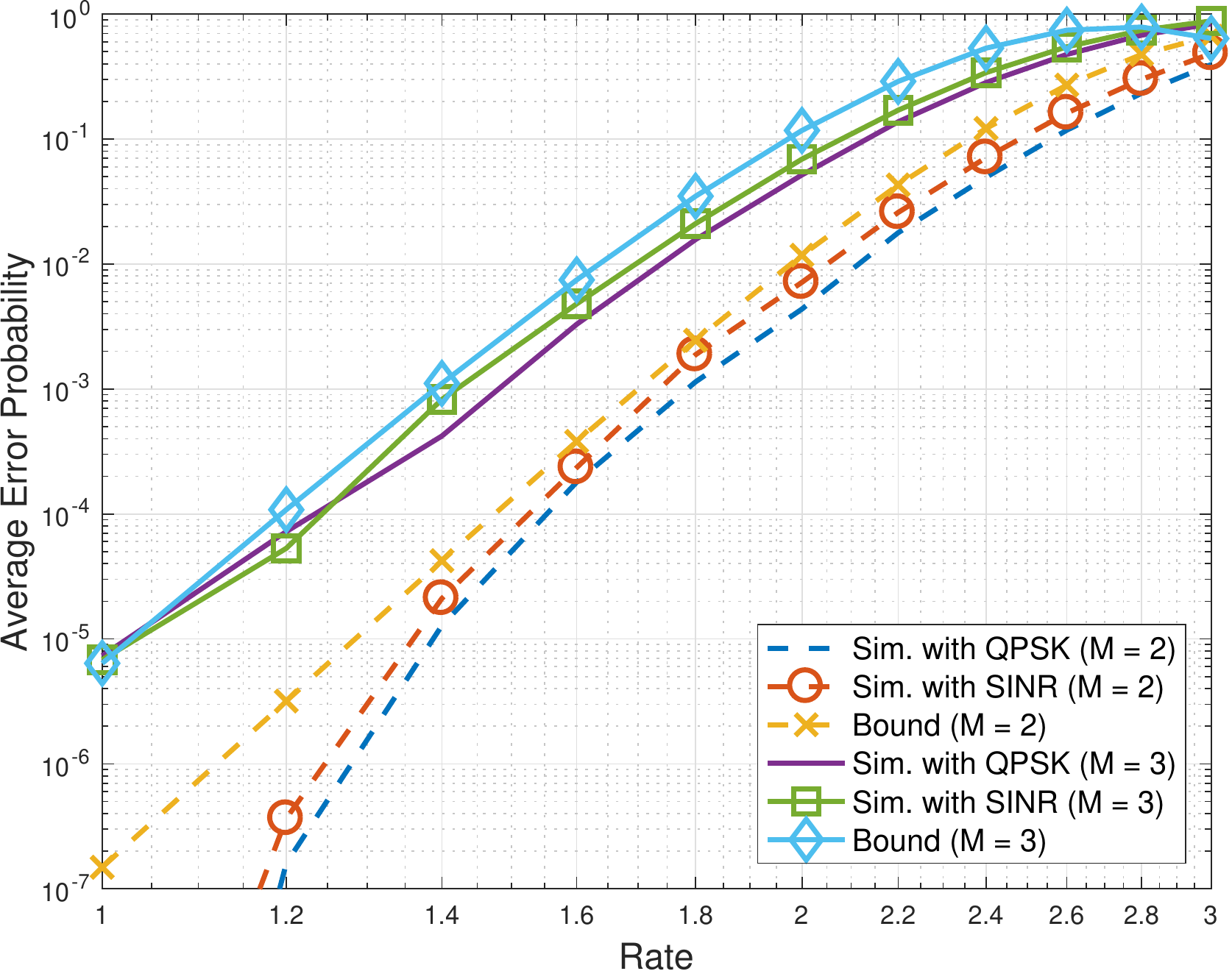} 
\end{center}
\caption{Average error probabilities as a function of $R$
with $D = 16$, $n = 512$, $\sSNR = 6$ dB, and $M \in \{2,3\}$.}
       \label{Fig:Eplt2}
\end{figure}

In Figs.~\ref{Fig:plt1} - ~\ref{Fig:plt_RvsER},
we consider the instantaneous SINR in \eqref{EQ:iSINR}
for simulations. 
Since \eqref{EQ:iSINR} is obtained under the assumption of {\bf A1},
it would be necessary to 
consider simulations with interleaved finite-length blocks.
To this end, quadrature phase shift keying
(QPSK) is considered with a block length of $n/2$
(since one QPSK symbol can transmit 2 bits).
Random interleaving at symbol-level is considered.
In Fig.~\ref{Fig:Eplt2},
we show the average error probability
for different values of the code rate
when there are $M \in \{2,3\}$ interfering signals,
$D = 16$, and $\sSNR = 6$ dB.
It is shown that the average error probability
with QPSK is slightly lower than that
with the instantaneous SINR in \eqref{EQ:iSINR}.
This may result from the fact that the correlation 
cannot be zero by symbol-level random interleaving 
and the correlation reduces the interference level.

\begin{figure}[thb]
\begin{center}
\includegraphics[width=\figwidth]{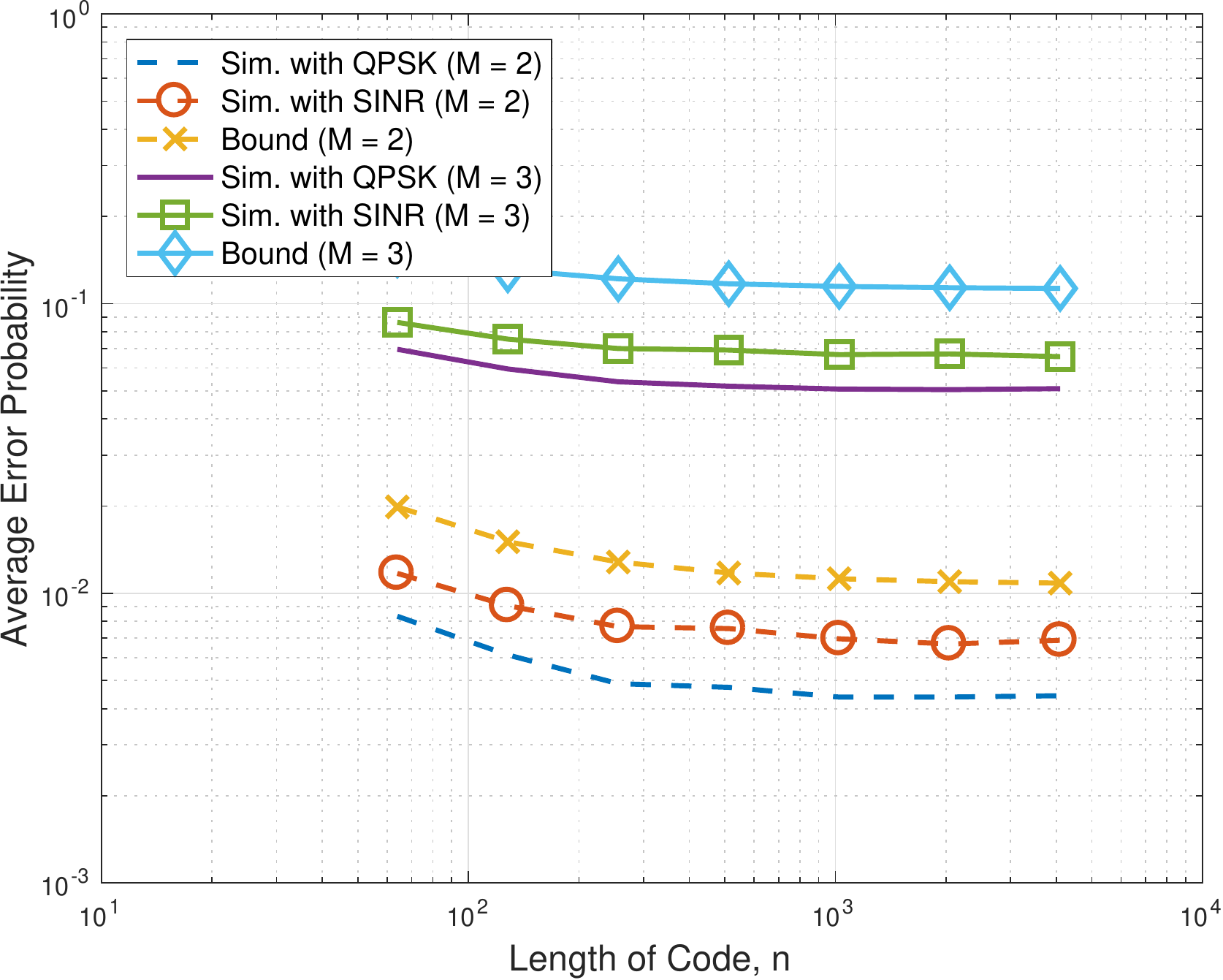} 
\end{center}
\caption{Outage probabilities as a function of $n$
with $D = 16$, $M = 2$, $R = 2$, and $\sSNR = 6$ dB.}
       \label{Fig:Eplt3}
\end{figure}

To see the impact of the length of finite-length codes,
$n$, on the average error probability,
we perform simulations and show the results in
Fig.~\ref{Fig:Eplt3} when
with $D = 16$, $M = 2$, $R = 2$, and $\sSNR = 6$ dB.
As expected, the average error probability
decreases with $n$, while it becomes saturated for a sufficiently
large $n$.
We can also confirm that 
the bound in \eqref{EQ:beps}
can be used to predict the performance with
finite-length codes from Figs.~\ref{Fig:Eplt2} and \ref{Fig:Eplt3}.

\section{Concluding Remarks}	\label{S:Conc}

In this paper, we discussed a NOMA scheme 
based on repetition to exploit high diversity gains.
The resulting scheme, called 
repetition-based NOMA, was able to provide a low error probability
without instantaneous CSI-based
power allocation thanks to high diversity gains.
In order to guarantee a target performance,
a closed-form expression for an upper-bound on 
the outage probability was derived so that
key parameters (e.g., the code rate) 
can be decided accordingly.
The case of finite-length codes was also considered
with the average error probability.
Simulation results demonstrated that the derived
upper-bound is reasonably tight and can be used to decide
key parameters that meet a certain target performance.

Since we mainly focused on the performance analysis
to derive a closed-form expression for the outage probability
in terms of key parameters, we did not work on
other issues, e.g., scheduler design.
The design of scheduler 
would be an interesting topic to be studied
in the future, which might be
based on the derived closed-form
expression for the outage probability in this paper.

\appendices

\section{Proof of Lemma~\ref{L:1}}	\label{A:1}

Since
$\uE[\chi_{2n}^2] = 2n$
and ${\rm Var}(\chi_{2n}^2) = 4n$,
it can be shown that
\begin{align}
\uE[\Omega^2] = \frac{1}{N^2} \left( 4 M^2 N^2 + 4N M \right) 
= 4 M \left( M+ \frac{1}{N}\right).
	\label{EQ:B2}
\end{align}

The 2nd moment of $W$ is given by
\begin{align}
\uE[W^2] & = \uE \left[ \sum_{l=1}^D \sum_{l^\prime = 1}^D 
\alpha_l \alpha_{l^\prime} Y_l Y_{l^\prime} \right]\cr
& = \sum_{l=1}^D \uE[\alpha_l^2] 
\uE[Y_l^2] + \sum_{l \ne l^\prime} \uE[\alpha_l\alpha_{l^\prime}] 
\uE[Y_l] \uE[Y_{l^\prime}]  \cr
& = D \alpha_{(2)} (4 M + 4M^2) 
+ 4 M^2 D (D-1) \sigma_{1,2},
	\label{EQ:W2o}
\end{align}
where $\alpha_{(2)} = \uE[\alpha_l^2]$
and $\sigma_{1,2} = \uE[\alpha_1 \alpha_2]$.
Since $X_{l,k}$ is an exponential random variable
under the assumption of {\bf A2},
$\alpha_l$ is expressed as
$\alpha_l = \frac{D_l}{\sum_{l^\prime=1}^D D_{l^\prime}}$,
where $D_l \sim {\rm Exp}(1)$ is an
independent exponential random variable with parameter 1.
The distribution of $\alpha_l$ is the same
as that of the minimum of 
$D-1$
independent standard uniform random variables \cite[Example 4.6]{Ahsanullah13},
i.e.,
$f(\alpha_l)  = (D-1) (1-\alpha_l)^{D-2}$, $\alpha_l \in [0,1]$. 
Thus, we have
\be
\alpha_{(2)} = \frac{2}{D (D+1)}.
	\label{EQ:a2D}
\ee
In addition, the distribution of 
$\alpha_1 + \alpha_2 = \frac{D_1 + D_2}{\sum_{l=1}^D D_l}$ 
is the same as the distribution of the 2nd smallest
order statistic among $D-1$
independent standard uniform random variables.
From this, since
$$
\uE[(\alpha_1 + \alpha_2)^2] = \uE[\alpha_1^2]
+ \uE[\alpha_2^2] + 2 \uE[\alpha_1 \alpha_2],
$$
we have
\begin{align}
\uE[\alpha_1 \alpha_2] 
& = \frac{\uE[(\alpha_1 + \alpha_2)^2]}{2} - \uE[\alpha_1^2] \cr
& = \frac{6}{2 D (D+1)} - \frac{2}{D (D+1)} 
= \frac{1}{D (D+1)},
\end{align}
because the 2nd moment of 
the 2nd smallest order statistic is $\frac{6}{D(D+1)}$ 
\cite[Eq. (8.4)]{Ahsanullah13}.
Then, it can be shown that
\be
\uE[W^2] = 
\frac{2}{D+1} (4 M + 4M^2) + \frac{D-1}{D+1} 4 M^2.
	\label{EQ:W2}
\ee
Consequently, from \eqref{EQ:B2} and \eqref{EQ:W2} we can find that
the 2nd moments of $W$ and $\Omega$ are the same if
$N$ is given as in \eqref{EQ:D1}, which completes the proof.

\section{Proof of Lemma~\ref{L:2}}	\label{A:2}

Using the Chernoff bound \cite{Mitz05},
it can be shown that
\begin{align}
\Pr(Z_D < z)
& \le  \uE[ e^{-t (Z_D - z)}] \cr
& = e^{2 D tz} \left( \frac{1}{1+ 2 t} \right)^D 
= \left(  \frac{e^{2t z}}{1+ 2 t} \right)^D.
\end{align}
Here, $t > 0$. Letting
$z = \frac{1}{1 + 2t}$,
we have
\be
\Pr(Z_D < z) \le  (z e^{1-z} )^D, \ z \in [0, 1),
        \label{EQ:CB_l}
\ee
which is reasonably tight.
For $z > 1$, it can also be shown that
\be
\Pr(Z_D > z) \le (z e^{1-z} )^D.
        \label{EQ:CB_r}
\ee

As in \cite{Choi20_WCNC}, the upper-bound
in \eqref{EQ:CB_l} can be tighter using 
a correction term as follows:
\begin{align}
\Pr(Z_D \le z) & \le F_D (z) \cr
& = (z c_D e^{1-z c_D } )^D, \ z \in [0, 1/c_D],
	\label{EQ:UB}
\end{align}
where $c_D$ is the correction term\footnote{\eqref{EQ:UB}  
with the correction term in \eqref{EQ:c_D} is 
an inequality conjecture.}
that is given in \eqref{EQ:c_D}.
Thus, if $M = 0$, \eqref{EQ:UB} can be applied to 
\eqref{EQ:OP_M0}, which results in
\eqref{EQ:M0}.

For the case that $M > 0$, we need to take into account
the interference.
In \eqref{EQ:app},
let $Y = \frac{\chi_{2 N M}^2}{2}$.
Then, using \eqref{EQ:UB}, it can be shown that
\begin{align}
\tilde \uP_k
& = 
\int_0^\infty \Pr\left(Z_D < \frac{T}{D} 
\left(\frac{y}{N} + \frac{1}{\sSNR} \right)
\right) f_Y (y) d y \cr
& \le \int_0^\kappa
( c_D \phi (y) e^{1- c_D \phi (y)} )^D f_Y (y) d y 
+ \int_\kappa^\infty
f_Y (y) d y \cr
& \le \int_0^\infty
( c_D \phi (y) e^{1- c_D \phi (y)} )^D f_Y (y) d y 
+ \int_\kappa^\infty
f_Y (y) d y , \quad 
	\label{EQ:UBk}
\end{align}
where $f_Y (y)$ represents the pdf of $Y$ and
\begin{align*}
\phi (y) = \frac{T}{D}
\left( \frac{y}{N} + \frac{1}{\sSNR} \right) \ \mbox{and} \
\kappa = N \left(\frac{D}{c_D T} - \frac{1}{\sSNR} \right).
\end{align*}
The first term on the RHS in  \eqref{EQ:UBk}
can be expressed as
\begin{align}
\psi & =
\left( \frac{c_D e T}{D} \right)^D 
e^{- \frac{c_D T}{\sSNR}} \cr
& \ \times \int_0^\infty
\left(\frac{y}{N} + \frac{1}{\sSNR}
\right)^D e^{- \frac{c_D T}{N} y}
f_Y( y) dy \cr
& = 
\frac{1}{D!} \left(\frac{T}{\sSNR} \right)^D 
e^{- \frac{c_D T}{\sSNR}} \cr
&\ \times 
\sum_{n = 0}^D \binom{D}{n} 
\left(\frac{\sSNR}{N}\right)^n
\int_0^\infty 
e^{-\frac{c_D T}{N} y}
\frac{ y^{M N-1+n} e^{-y}}{(M N-1)!} dy \cr
& = 
\frac{1}{D!} \left(\frac{T}{\sSNR} \right)^D 
e^{- \frac{c_D T}{\sSNR}} \cr
&\ \times 
 \sum_{n = 0}^D \binom{D}{n} \left( \frac{\sSNR}{N} \right)^{n} 
\left(\frac{1}{1+ \frac{c_D T}{N}
} \right)^{M N+n}
\frac{(M N +n-1)!}{(M N-1)!} \cr
& = 
\frac{1}{D!} \left(\frac{T}{\sSNR} \right)^D 
\frac{e^{- \frac{c_D T}{\sSNR}}}
{(1 + \frac{c_D T}{N})^{M N}}
\cr
&\ \times 
 \sum_{n = 0}^D \binom{D}{n} \left( \frac{\sSNR}{N
+ c_D T} \right)^{n} 
\frac{(M N +n-1)!}{(M N-1)!},
	\label{EQ:psi0}
\end{align}
which becomes \eqref{EQ:psi}.

From \eqref{EQ:CB_r},
the 2nd term on the RHS in  \eqref{EQ:UBk}
is bounded as follows:
\begin{align}
\Pr(Y \ge \kappa)
& = \Pr\left(
\frac{\chi_{2 N M}^2}{2 N M}
\ge \frac{1}{M} \left( \frac{D}{c_D T} - \frac{1}{\sSNR} \right)
\right) \cr
& \le 
\left( \frac{d}{M} e^{1 - \frac{d}{M} }
\right)^{NM} 
= \left(\frac{d e}{M} \right)^{NM} e^{-d N},
\end{align}
which
is the 2nd term on the RHS in \eqref{EQ:app2}.
This completes the proof.

\section{Proof of Lemma~\ref{L:x}}	\label{A:x}
Using 
the inequality of arithmetic and geometric means,
it can be shown that
\begin{align}
\prod_{t=0}^{n-1}(M N+t) 
\le
\left(MN + \frac{\sum_{t=0}^{n-1}t }{n} \right)^n 
= 
\left(MN + \frac{n-1}{2} \right)^n.
	\label{EQ:AGineq}
\end{align}
In \eqref{EQ:psi}, since $n \le D$,
using \eqref{EQ:AGineq},
it follows
\begin{align}
& \sum_{n=0}^D \binom{D}{n} 
\left(\frac{\sSNR}{N + c_D T} \right)^n
\prod_{t=0}^{n-1}(M N+t)  \cr
& \le 
\left( 1 + 
\frac{\sSNR}{N + c_D T}  \left(MN + \frac{D-1}{2} \right) \right)^D.
	\label{EQ:bb1}
\end{align}
Substituting \eqref{EQ:bb1}
into \eqref{EQ:psi}, with $N = \frac{D+1}{2}$, we have
\begin{align}
\psi & 
\le
\frac{e^{- \frac{c_D T}{\sSNR}}}{D!} 
\left(\frac{T}{\sSNR} \right)^D
\left(1+ \frac{c_D T}{N}\right)^{-M N}\cr
& \ \times
\left( 1 + 
\frac{\sSNR}{N + c_D T}  \left(MN + \frac{D-1}{2} \right) \right)^D \cr
& \le \frac{(1+ \frac{c_D T}{N})^{-\frac{M}{2}}}{D!}  \nu^D,
	\label{EQ:uu1}
\end{align}
where
$
\nu =
\frac{T}{\sSNR} \left(1 + \frac{2 c_D T}{D+1} \right)^{-\frac{M}{2}} 
\left( 1 + 
\frac{\sSNR(D(M+1) + M-1) }{D+1 + 2 c_D T}  
\right)$.
In \eqref{EQ:uu1}, the last inequality is due to the
fact that $e^{-\frac{c_D T}{\sSNR}} \le 1$.
Then, we can see that $\nu < 1$ if \eqref{EQ:condx} holds.

From \eqref{EQ:cond2}, we have
\be
c_D T = \frac{ D \sSNR}{d \sSNR + 1} \approx \frac{D}{d}, \ \sSNR \gg 1.
	\label{EQ:cDT}
\ee
To determine $C$ in 
\eqref{EQ:p_nu}, from \eqref{EQ:cDT}, it can be shown that
\begin{align}
\frac{1}{1 + \frac{c_D T}{N}}
& = \frac{1}{1 + \frac{2 c_D T}{D+1}}
= \frac{1}{1 + \frac{2 D \sSNR}{(D+1) (d \sSNR +1)}} \cr
& \le \frac{1}{1 + \frac{\sSNR}{d \sSNR +1}}
= \frac{1 + d \sSNR}{1+(1+d) \sSNR},
\end{align}
we have
$C = \left(\frac{1}{1 + \frac{\sSNR}{d \sSNR +1}} \right)^{\frac{M}{2}}
\ge \left( \frac{1}{1+ \frac{c_D T}{N}}\right)^{\frac{M}{2}}$,
where $C$ is independent of $D$.

\section{Proof of Lemma~\ref{L:3}}	\label{A:3}

As $\sSNR \to \infty$, from \eqref{EQ:psi0},
after some manipulations, we can show that
\begin{align}
\bar \psi & = 
\left( \frac{c_D e T}{D} \right)^D 
\int_0^\infty
\left(\frac{y}{N} \right)^D e^{- \frac{c_D T}{N} y}
f_Y( y) dy \cr
& = 
\left( \frac{c_D e T}{D N} \right)^D 
\frac{\Gamma (MN + D)}{\Gamma (MN)} 
\frac{1}{\left( 1 + \frac{c_D T}{N} \right)^{MN + D}}\cr
& = 
\frac{1}{D!} 
\frac{\Gamma (MN + D)}{\Gamma (MN)} 
\left(\frac{T}{N+c_D T} \right)^D 
\frac{1}{\left( 1 + \frac{c_D T}{N} \right)^{MN}}\cr
& = \binom{MN+D-1}{MN-1}
\left(\frac{T}{N+c_D T} \right)^D 
\left(\frac{N}{ N + c_D T}\right)^{MN},
\end{align}
which becomes \eqref{EQ:L3}.

\bibliographystyle{ieeetr}
\bibliography{noma}
\end{document}